\documentstyle[preprint,aps]{revtex}

\def\coeff#1#2{{\textstyle{#1\over #2}}}
\def\half{\coeff{1}{2}}

\def\pmb#1{\setbox0=\hbox{$#1$}%
\kern-.025em\copy0\kern-\wd0
\kern.05em\copy0\kern-\wd0
\kern-.025em\raise.0433em\box0}

\def\beq{\begin{equation}}
\def\eeq{\end{equation}}

\def\yangmills{\vcenter{\hbox to 48pt{\hss\vbox to 20pt{\vss
\vss}\hss}}}

\def\ymdash{\vcenter{\hbox to 48pt{\hss\vbox to 20pt{\vss
\vss}\hss}}}

\def\ymobd{\vcenter{\hbox to 48pt{\hss\vbox to 20pt{\vss
\vss}\hss}}}

\def\ymtbd{\vcenter{\hbox to 48pt{\hss\vbox to 20pt{\vss
\vss}\hss}}}

\def\ymloop{\vcenter{\hbox to 0.3in{\hss\vbox to 0.4in{\vss
\vss}\hss}}}

\def\ymloopbd{\vcenter{\hbox to 0.3in{\hss\vbox to 0.4in{\vss
\vss}\hss}}}

\def\ymblip{\vcenter{\hbox to 24pt{\hss\vbox to 0.4in{\vss
\vss}\hss}}}

\def\section{{\setcounter{equation}{0}}
\@mainheadtrue
\@startsection {section}{1}{\z@}{0.8cm plus1ex minus
 .2ex}{0.5cm plus1ex minus.2ex}{\reset@font\small\bf\centering}
           }
\makeatother

\begin{document}

\def\footnoterule{\hrule width \hsize}
\skip\footins = 14pt 
\footskip     = 20pt 
\footnotesep  = 12pt 


\textwidth=6.5in
\hsize=6.5in
\oddsidemargin=0in
\evensidemargin=0in
\hoffset=0in

\textheight=9.5in
\vsize=9.5in
\topmargin=-.5in
\voffset=-.3in

\title{%
Threshhold Singularities and the Magnetic Mass in Hot QCD%
\footnotemark[1]}

\def\footstrut{\baselineskip 16pt}

\footnotetext[1]
{\baselineskip=16pt
This work is supported in part
by funds provided by
the U.S.~Department of Energy (D.O.E.)
under contracts \#DE-FC02-94ER40818 and \#DE-FG02-91ER40676.
\hfil\break
BU-HEP-95-31, MIT-CTP-2479 \hfil
October 1995\break
}

\author{R. Jackiw}

\address{Center for Theoretical Physics,
Laboratory for Nuclear Science
and Department of Physics \\[-1ex]
Massachusetts Institute of Technology,
Cambridge, MA ~02139--4307}

\author{So-Young Pi}

\address{Department of Physics,
Boston University,
Boston, MA ~02215}

\maketitle

\setcounter{page}{0}
\thispagestyle{empty}

\begin{abstract}%
\noindent%
\baselineskip=16pt%

We discuss the recently devised one-loop gap equation for the magnetic
mass of hot QCD.  An alternative, and one would hope equivalent, gap
equation is presented, which however shows no mass generation at the
one-loop level.

\end{abstract}

\vspace*{\fill}
\begin{center}
Submitted to {\it Physics Letters B}
\end{center}
\vspace*{\fill}

\newpage

\baselineskip=24pt plus 1pt minus 1pt

\widetext

In an Abelian plasma, static electric fields are screened (Debye mass or
screening length); there is no magnetic screening since there are no
magnetic sources.  When this problem is treated by thermal quantum field
theory, the electric screening mass straightforwardly emerges from
Feynman diagrams at high-temperature $T$, and is found to be of order
$eT$, where $e$ is the coupling strength.  In a resummed perturbation
expansion, this mass also cures the infrared divergences that afflict
un-resummed finite-temperature perturbative expansions when there are
massless fields in the theory.

Similar electric mass generation has been demonstrated for non-Abelian
gauge theories, but this does not remove all the infrared divergences,
which remain when the non-linear interactions of electric (temporal) and
magnetic (spatial) degrees of freedom are treated perturbatively
\cite{ref:1}.  While it is believed that these divergences are also
cured by the generation of a magnetic mass $\mu$, a convincing
calculation for $\mu$ is thus far unavailable.  The perturbative
resummation, which exposes the Debye mass, gives no evidence for a
magnetic mass.

A similar problem arises in three-dimensional (Euclidean) Yang-Mills
theory at {\it zero\/} temperature, which should provide an effective
description for the magnetic (spatial) degrees of freedom of
four-dimensional QCD at high-temperature, through the identification of
the three dimensional coupling $g$ with $e \sqrt{T}$.  Since $g^2$
carries dimension of mass, it is plausible to suppose that
three-dimensional Yang-Mills theory dynamically generates an
$O(g^2)$ mass, which eliminates perturbative infrared divergences in the
three-dimensional model, and suggests the occurrence of an $O(e^2T)$
magnetic mass in the four-dimensional theory.  However, thus far no
analysis of the three-dimensional Yang-Mills model has led to a proof of
mass generation.

Since the mass is not seen in perturbative expansions, even resummed
ones, one attempts a non-perturbative calculation, based on a gap
equation.  Of course an exact treatment is impossible; one must be
satisfied with an approximate gap equation, which effectively sums a
large, but still incomplete set of graphs.  At the same time, gauge
invariance should be maintained; gauge non-invariant approximations are
not persuasive.

Deriving an approximate, but gauge invariant gap equation is most
efficiently carried out in a functional integral formulation.  We begin
by reviewing how a one-loop gap equation is gotten from the functional
integral, first for a non-gauge theory of a scalar field $\varphi$,
then we indicate how to extend the procedure when gauge invariance is to
be maintained for a gauge field $A_\mu$.

Consider a self-interacting scalar field theory
(in the Euclidean formulation)
whose potential $V(\varphi)$ has no quadratic term,
so in direct perturbation theory one
may encounter infrared divergences, and one enquires whether a mass is
generated, which would cure them.
\begin{eqnarray}%
{\cal L} &=& \half \partial_\mu \varphi \partial^\mu \varphi +
V(\varphi) \nonumber\\
V(\varphi) &=& \lambda_3 \varphi^3 + \lambda_4 \varphi^4 + \ldots
\label{eq:1}
\end{eqnarray}%
The functional integral involves
the negative exponential of the action $I = \int {\cal L}$.   Separating
the quadratic, kinetic part of $I$, and expanding the exponential of the
remainder in powers of the field yields the usual loop expansion, which
may also be systematized by introducing a loop-counting parameter $\ell$
and considering
$e^{-{1\over\ell} I (\sqrt{\ell} \varphi)}$:  the power series in $\ell$
is the loop expansion.  To obtain a gap equation for a possible mass
$\mu$, we proceed by adding and subtracting $I_\mu = {\mu^2 \over 2}
\int \varphi^2$, which of course changes nothing.
\beq
I = I + I_\mu - I_\mu
\label{eq:2}
\eeq
Next the loop expansion is reorganized by expanding $I + I_\mu$ in the
usual way, but taking $-I_\mu$ as contributing    at one    loop higher.
This is systematized by replacing (\ref{eq:2}) with an effective action,
$I_\ell$,
containing the loop counting  parameter $\ell$, which organizes    the
loop expansion    in the   indicated manner.
\beq
I_\ell = {1\over\ell} \left( I (\sqrt{\ell} \varphi)
+ I_\mu (\sqrt{\ell} \varphi) \right) - I_{\mu} (\sqrt{\ell} \varphi)
\eeq
An expansion in powers of $\ell$ corresponds to a resummed series;
keeping all terms and setting $\ell$ to unity returns us to the original
theory (assuming that rearranging the series does no harm);
approximations consist of keeping a finite number of terms: the
$O(\ell)$ term involves a single loop.

The gap equation is gotten by considering
the self energy $\Sigma$ of the complete propagator.  To one-loop order,
the contributing graphs are depicted in Fig.~1.

\beq
\Sigma = {~}
\hbox to 0pt{\kern0pt\lower8pt\hbox{$\scriptstyle{\lambda_3}$}}
\yangmills
\hbox to 0pt{\kern-7pt\lower8pt\hbox{$\scriptstyle{\lambda_3}$}}
{~} +
\hbox to 0pt{\kern8pt\lower16pt\hbox{$\scriptstyle{\lambda_4}$}}
\ymloop
-
\hbox to 0pt{\kern8pt\lower16pt\hbox{$\scriptstyle{\mu^2}$}}
\ymblip
\eeq
\centerline{{\rm Fig.~1.~~Self energy to resummed one-loop order.}}
\smallskip

Regardless of the form of the exact potential, only the three- and four-
point vertices are needed at one-loop order; the bare propagators are
massive thanks to the addition of the mass term ${1\over \ell} I_\mu
(\sqrt{\ell} \varphi) = {\mu^2 \over 2} \int \varphi^2$; the last $-\mu^2$
in Fig.~1 comes from the subtraction of the same mass term,
but at one-loop order: $-I_\mu (\sqrt{\ell} \varphi) = - \ell {\mu^2 \over
2} \int \varphi^2$.

The gap equation emerges when it is demanded
that $\Sigma$ does not shift the mass $\mu$.
In momentum space, we require
\begin{eqnarray}
\Sigma(p) \biggr|_{p^2 = -\mu^2} &=& 0 \\[1.0ex]
\hbox to 0pt{\kern0pt\lower8pt\hbox{$\scriptstyle{\lambda_3}$}}
\yangmills
\hbox to 0pt{\kern-7pt\lower8pt\hbox{$\scriptstyle{\lambda_3}$}}
{~} +
\hbox to 0pt{\kern8pt\lower16pt\hbox{$\scriptstyle{\lambda_4}$}}
\ymloop
{~} \Biggr|_{p^2 = -\mu^2} &=& \mu^2
\nonumber
\end{eqnarray}%
\centerline{{\rm Fig.~2.~~Graphical depiction of Eq.~(5).}}
\smallskip

While the above ideas can be applied to a gauge theory, it is
necessary to elaborate them so that gauge invariance is preserved.  We
shall discuss solely the three-dimensional non-Abelian Yang-Mills model (in
Euclidean formulation) as an interesting theory in its own right, and
also as a key to the behavior of spatial variables in the physical,
four-dimensional model at high temperature.

The starting action $I$ is the usual one for a gauge field.
\begin{eqnarray}
I &=& \int d^3 x {\rm ~tr~} {\half} F^{i} F^{i} \nonumber\\
F^{i} &=& {\half} \epsilon^{ijk} F_{jk}
\label{eq:4}
\end{eqnarray}
While one may still add and subtract a mass-generating term $I_\mu$, it is
necessary to preserve gauge invariance.  Thus we seek a gauge invariant
functional of $A_i$, $I_\mu(A)$, whose quadratic portion
gives rise to a mass.  Evidently
\beq
I_\mu (A) = -{\mu^2 \over 2} \int d^3 x {\rm ~tr~} A_i
\left( \delta_{ij} - {\partial_i \partial_j \over \nabla^2} \right)
A_j + \ldots
\label{eq:7}
\eeq
The transverse structure of (\ref{eq:7})
guarantees invariance against
Abelian gauge transformations;
the question then remains how the quadratic term is to be completed in
order that $I_\mu(A)$ be invariant
against non-Abelian gauge transformations.
[In fact for the one-loop gap
equation only terms through $O(A^4)$ are needed.]

A very interesting proposal for $I_\mu(A)$ was given by Nair
\cite{ref:1,ref:2} who also put forward the scheme for determining the
magnetic mass, which we have been describing.  By modifying in various
ways the hard thermal loop generating functional (which gives a
four-dimensional, gauge invariant but Lorentz non-invariant effective
action with a transverse quadratic term), he arrived at a gauge and
rotation invariant three-dimensional structure, which can be employed in
the derivation of a gap equation.

The scheme proceeds as in the scalar theory,
except that $I_\mu(A)$ gives rise not only to a mass term for the free
propagator,
but also to higher-point interaction vertices.
At one loop only the three- and four- point vertices are needed, and
to this order the subtracting term uses only the
quadratic contribution.  Thus the gap equation reads
\beq
\left[ {~}
\ymdash ~+~ \yangmills ~+ \ymloop  + ~\ymobd~ + ~\ymtbd~ + \ymloopbd
\right]_{p^2 = -\mu^2} =
{~}
\hbox to 0pt{\kern8pt\lower16pt\hbox{$\scriptstyle{\mu^2}$}} \ymblip
\eeq
\centerline{{\rm Fig.~3.~~Graphical depiction of Yang-Mills gap equation.}}
\smallskip

The first three graphs are as in ordinary Yang-Mills theory, with
conventional vertices, but massive gauge field
propagator (solid line);
\beq
D_{ij} (p) = \delta_{ij} {1\over p^2 + \mu^2}
\label{eq:5}
\eeq
the first graph depicting the gauge compensating ``ghost'' contribution,
has massless ghost propagators
(dotted line)
and vertices determined by the
quantization gauge, conveniently chosen, consistent with (\ref{eq:5}),
to be
\beq
{\cal L}_{{\rm gauge}\atop{\rm fixing}}
= {\textstyle{1\over2}}
\pmb{\nabla} \cdot {\bf A}
(1-\mu^2/\nabla^2)
\pmb{\nabla} \cdot {\bf A}
\label{eq:6}
\eeq
The remaining three graphs arise from Nair's form for
hard thermal loop-inspired
$I_\mu(A)$, with
solid circles denoting the new vertices.  As it happens, the last graph
with the four-point vertex vanishes, while the three-point vertex reads
\begin{eqnarray}
&& \hskip-4pt
{}^N \! V^{abc}_{ijk} ({\bf p}, {\bf q}, {\bf r}) =
\nonumber\\
&& \hskip-4pt
- {i \mu^2 \, f^{abc} \over 3! ({\bf p} \times {\bf q})^2}
\,
\left\{ {1\over3}
\left(
{{\bf p} \cdot {\bf q} \over p^2} + {{\bf r} \cdot {\bf q} \over r^2}
\right)
p_i p_j p_k
- {{\bf r} \cdot {\bf p} \over 3 r^2}
(q_i q_j p_k + q_i p_j q_k + p_i q_j q_k )
\right\}
+ {\scriptstyle{\rm 5~permutations}}
\label{eq:11}
\end{eqnarray}
\centerline{$p+q+r=0$}

The permutations ensure that the vertex is symmetric under the exchange
of any pair of index sets $(a~i~p), (b~j~q), (c~k~r)$.
[We discuss the $SU(N)$ theory, with structure constants $f^{abc}$.]

The result of the computation is
\begin{mathletters}%
\begin{eqnarray}
{}^N \! \Pi^{ab}_{ij} &=& \delta^{ab} \Pi_{ij}^N \\
\Pi_{ij}^N &=& \Pi_{ij}^{YM} + \overline{\Pi}_{ij}^N
\end{eqnarray}
\end{mathletters}
$\Pi_{ij}^{YM}$ is the contribution from the first three Yang-Mills
graphs and $\overline{\Pi}_{ij}^N$ sums the graphs from $I_\mu(A)$.
The reported results are \cite{ref:3}
\begin{eqnarray}
\hspace*{-.4in}
\Pi_{ij}^{YM} (p) &=& N (\delta_{ij} - \hat{p}_i \hat{p}_j)
\left[
\left( {- 13 p^2 \over 64 \pi \mu} + {5 \mu \over 16\pi} \right)
{2\mu \over p} \tan^{-1} {p \over 2\mu} - {\mu \over 16\pi} - {p \over 64}
\right] \nonumber\\
&& \hbox{\qquad} + N \hat{p}_i \hat{p}_j
\left[
\left(
{p^2 \over 32 \pi \mu} + {\mu \over 8\pi} \right)
{2\mu \over p} \tan^{-1}{p \over 2\mu} + {\mu \over 8\pi} - {p \over 32}
\right] \label{eq:013} \\
\overline{\Pi}_{ij}^N (p)
&=& N \left( \delta_{ij} - \hat{p}_i \hat{p}_j \right)
\left[
\left( {3 p^2 \over 64 \pi \mu} + {3\mu \over 16\pi} \right)
{2\mu \over p} \tan^{-1} {p \over 2\mu}
- {p^2 \over 8\pi \mu}
\left( {\mu^2 \over p^2} + 1 \right)^2
{\mu \over p} \tan^{-1} {p \over \mu}
+ {\mu \over 16\pi} + {\mu^3 \over 8\pi p^2} + {p \over 64}
\right]
\nonumber\\
&& \hbox{\qquad} - N \hat{p}_i \hat{p}_j
\left[
\left(
{p^2 \over 32 \pi \mu} + {\mu \over 8\pi} \right)
{2\mu \over p} \tan^{-1} {p \over 2\mu} +
{\mu \over 8\pi} - {p \over 32} \right]
\label{eq:9}
\end{eqnarray}
The Yang-Mills contribution (\ref{eq:013})
is not separately gauge-invariant (transverse) owing to the massive
gauge propagators.
[At $\mu = 0$, $\Pi^{YM}_{ij}$ reduces to the standard result \cite{ref:4}:
$N (\delta_{ij} - \hat{p}_i \hat{p}_j ) \left( - {7 \over 32} p \right)$.]
The longitudinal terms in
$\Pi^{YM}_{ij}$ are cancelled by those in
$\overline{\Pi}^N_{ij}$, so that the total is transverse.
\beq
\Pi_{ij}^N (p) = N \left( \delta_{ij} - \hat{p}_i \hat{p}_j \right)
\left[
\left(
{- 5 p^2 \over 32 \pi \mu} + {1 \over 2\pi} \mu \right)
{2 \mu \over p} \tan^{-1} {p \over 2\mu}
- {p^2 \over 8 \pi \mu} \left(  {\mu^2 \over p^2} + 1 \right)^2
{\mu \over p} \tan^{-1} {p \over \mu} + {\mu^3 \over 8 \pi p^2} \right]
\label{eq:015}
\eeq
[Dimensional regularization is used to avoid divergences.]

Before proceeding, let us note the analytic structures in the above
expressions, which are presented for Euclidean momenta, but have to be
evaluated at the Minkowski value $p^2 = - \mu^2 < 0$.
Analytic continuation for the inverse tangent is provided by
\beq
{1\over x} \tan^{-1} x = {1 \over 2 \sqrt{-x^2}}
\, \ln \, {1 + \sqrt{-x^2} \over 1 - \sqrt{-x^2}}
\label{eq:16}
\eeq
Evidently $\Pi_{ij}^N(p)$ possesses threshhold singularities,
at various values of $-p^2$.

There is a singularity at $p^2 = -4 \mu^2$
(from $\tan^{-1} {p\over 2\mu}$)
arising because the graphs in Fig.~3, containing massive propagators
(\ref{eq:5}), describe the excahnge of two massive gauge ``particles''.
Moreover, there is singularity at $p^2 = -\mu^2$
(from $\tan^{-1} {p\over\mu}$)
and also, separately in $\Pi_{ij}^{YM}$ and $\overline{\Pi}^N_{ij}$,
at $p^2 = 0$ (from the $\pm {p \over 64}, \pm {p \over 32}$ terms).
These are understood in the following way.  Even though the propagators
are massive, the three-point function (\ref{eq:11}) contains
${1\over p^2}$,
${1\over q^2}$,
${1\over r^2}$
contributions, which act like massless propagators.  Thus the
threshhold at $p^2 = -\mu^2$ arises from the exchange of a massive line
(propagator) together with a massless line (from the vertex).  Similarly
the threshhold at $p^2 = 0$ arises from the massless lines in the vertex
(and also from massless ghost exchange).
The expressions acquire an imaginary part when the largest threshhold,
$p^2=0$, is crossed:  $\Pi_{ij}^{YM}$ and $\overline{\Pi}_{ij}^N$
are complex for $p^2 < 0$.

In the complete answer, the $p^2 = 0$ threshholds cancel, and the
singularity at the $p^2 = -\mu^2$ threshhold is extinguished by the
factor $({\mu^2 \over p^2} + 1)^2$.
Consequently
$\Pi_{ij}^N$
becomes complex only for $p^2 < -\mu^2$, and is real, finite at
$p^2 = -\mu^2$.
\beq
\Pi^N_{ij} (p) \biggr|_{p^2 = -\mu^2} =
(\delta_{ij} - \hat{p}_i \hat{p}_j)
\, {N\mu \over 32\pi} \,
(21 \ln 3 - 4)
\label{eq:13}
\eeq

{}From the gap equation in Fig.~3, the result for the mass is \cite{ref:3}
\beq
\mu = {N \over 32\pi} \, (21 \ln 3 - 4) \sim  {2.384 N \over 4\pi}
\label{eq:14}
\eeq
[in units of the coupling constant $g^2$ (or $e^2 T$), which has been
scaled to unity].

Before accepting the plausible answer (\ref{eq:14}) for $\mu$, it is
desirable     to assess higher order corrections, for example two-loop
contributions.  Unfortunately, an estimate \cite{ref:3} indicates that
79 graphs have to be evaluated, and the task is formidable.

Here we propose an alternative test for the reliability of the above
approach and the stability of the result (\ref{eq:14}) against
corrections.

We suggest deriving the gap equation with a different gauge invariant
completion to (\ref{eq:4}).  Rather than taking inspiration from hard
thermal loops (which after all have no intrinsic relevance to the
three-dimensional gauge theory\footnote{The hot thermal loop generating
functional is related to the Chern-Simons eikonal, see Ref.~[1].
Since the Chern-Simons term {\it is\/} a three-dimensional structure,
this fact may provide a basis for establishing the relevance of the hard
thermal loop generating functional to three-dimensional Yang Mills
theory.  The point is under investigation by D.~Karabali and V.~P.~Nair
(private communication).}), we take the following formula for
$I_\mu$,
\beq
I_\mu(A) = {\mu^2} \int d^3 x {\rm ~tr~} F^{i} {1\over D^2} F^{i}
\label{eq:15}
\eeq
where $D^2$ is the gauge covariant Laplacian.  While ultimately there is
no {\it apriori\/} way to select one gauge-invariant completion to
(\ref{eq:4}) over another, we remark that expressions like (\ref{eq:14})
appear in two-dimensional gauge theories (Polyakov gravity action,
Schwinger model) and are responsible for mass generation.
If two- and higher- loop effects are indeed ignorable, this alternative
gauge invariant completion, which corresponds to an alternative
resummation, should produce an answer close to (\ref{eq:14}).

With (\ref{eq:15}), the graphs are again as in Fig.~3, where the
propagator is still given by (\ref{eq:5}) in the gauge (\ref{eq:6}).
However, the three- and four- point vertices in $I_\mu(A)$ are different.
One now finds for the three-point vertex
\beq
V_{ijk}^{abc} ({\bf p},{\bf q},{\bf r}) = {-i \mu^2 \over 3!}
f^{abc}
\left(
\delta_{ij}
{\bf q} \cdot {\bf r} + q_i p_j
\right)
{p_k \over p^2 q^2}
+ {\scriptstyle {\rm ~5~permutations}}
\label{eq:016}
\eeq
\centerline{$p+q+r=0$}

and the four-point vertex reads
\begin{eqnarray}
&& \hspace*{-.3in}
V^{abcd}_{ijkl} ({\bf p},{\bf q},{\bf r},{\bf s}) =
{- \mu^2 \over 4!} f^{abe} f^{cde}
\left\{
{1\over2} \delta_{jk} \, \epsilon_{imn} \, \epsilon_{\ell on} \,
{p_m \over p^2} \, {s_0 \over s^2}
\right.
\\
&& - {1 \over 2 r^2}
\left.
\left(
{1\over4} \epsilon_{ijm} \epsilon_{k\ell m} -
\epsilon_{imn} \epsilon_{k\ell n}
{p_m \over p^2}
(p-r-s)_{j}
+ \epsilon_{imn} \epsilon_{\ell on} {p_m \over p^2} {s_0 \over s^2}
(p-r-s)_j (p+q-s)_k
\right)
\right\} \nonumber\\
&&
+ {\scriptstyle{\rm ~23~permutations}}
\nonumber
\end{eqnarray}
\centerline{$p+q+r+s=0$}

With these, the last three graphs in Fig.~3 are evaluated with the help
of dimensional regularization, and one finds
\begin{eqnarray}
&& \hspace*{-.3in}
\overline{\Pi}_{ij} (p) = N (\delta_{ij} - \hat{p}_i \hat{p}_j)
\left(
\left(
{p^6 \over 128\pi \mu^5}
+ {p^4 \over 32\pi \mu^3}
+ {7p^2 \over 64\pi \mu}
+ {27\mu \over 64\pi}
- {\mu^3 \over 16\pi p^2} \right)
{2\mu\over p} \tan^{-1} {p \over 2\mu}
\right.
\nonumber\\
&&
\left.
-
\left(
{p^6 \over 32\pi \mu^5}
+ {p^4 \over 16 \pi \mu^3}
- {p^2 \over 16 \pi \mu}
+ {\mu \over 32\pi} \right)
\left( {\mu^2 \over p^2} + 1 \right)^2
{\mu \over p} \tan^{-1} {p \over \mu}
\right.
\nonumber\\
&&
\left.
- {p^2 \over 32 \pi \mu}
- {3\mu \over 16 \pi}
+ {49 \mu^3 \over 96 \pi p^2}
+ {\mu^5 \over 32 \pi p^4}
+ {p^5 \over 128 \mu^4}
+ {p^3 \over 32 \mu^2}
- {p \over 16}
\right)
\nonumber\\
&&
- N \hat{p}_i \hat{p}_j
\left(
\left(
{p^2 \over 32 \pi \mu}
+ {\mu \over 8\pi}
\right)
{2\mu \over p}
\tan^{-1} {p \over 2\mu}
+ {\mu \over 8\pi}
- {p \over 32}
\right)
\label{eq:18}
\end{eqnarray}
A check on this very lengthy calculation is that summing it with
Yang-Mills contribution (\ref{eq:013}) yields a transverse result.
\begin{eqnarray}
&& \hspace*{-.3in}
\Pi_{ij}(p) = N (\delta_{ij} - \hat{p}_i \hat{p}_j)
\left(
\left(
{p^6 \over 128 \pi \mu^5}
+ {p^4 \over 32\pi \mu^3}
- {3 p^2 \over 32 \pi \mu}
+ {47 \mu \over 64\pi}
- {\mu^3 \over 16 \pi p^2} \right)
{2\mu \over p} \tan^{-1} {p \over 2\mu}
\right.
\nonumber\\
&&
\left.
- \left(
{p^6 \over 32 \pi \mu^5}
+ {p^4 \over 16\pi \mu^3}
- {p^2 \over 16 \pi \mu}
+ {\mu \over 32\pi}
\right)
\left( {\mu^2 \over p^2} + 1 \right)^2
{\mu \over p} \tan^{-1} {p \over \mu}
\right.
\nonumber\\
&&
\left.
- {p^2 \over 32 \pi \mu}
- {\mu \over 4\pi}
+ {49 \mu^3 \over 96 \pi p^2}
+ {\mu^5 \over 32 \pi p^4}
+ {p^5 \over 128 \mu^4}
+ {p^3 \over 32\mu^2}
- {5p \over 64}
\right)
\label{eq:19}
\end{eqnarray}
Another check on the powers of ${p \over \mu}$ is that the above reduces
to the Yang-Mills result at $\mu = 0$.

Just as (\ref{eq:013})--(\ref{eq:015}), the present formula exhibits threshhold
singularities: at $-p^2 = 4 \mu^2$, which are beyond our desired evaluation
point
$-p^2 = \mu^2$;  there are also threshhold singularities at $-p^2 = \mu^2$,
which are extinguished by the factor $({\mu^2 \over p^2} + 1)^2$; however,
those
at $p^2 = 0$ do {\it not\/} cancel, in contrast to the previous case --- indeed
$\Pi_{ij}(p)$ diverges at $p^2 = 0$, and is complex for $p^2 < 0$.
$\Bigl[$ It is interesting to remark that the last graph of Fig.~3, involving
the four-point vertex, which vanishes in the previous evaluation, here gives a
transverse result with unextinguished threshhold singularities at $-p^2 =
\mu^2$ and at $p^2 = 0$.  The protective factor of
$({\mu^2 \over p^2} + 1)^2$ arises when the remaining two graphs are added to
form $\overline{\Pi}_{ij}$ of (\ref{eq:18}),  and these also contain
non-cancelling $p^2 = 0$ threshhold singularities, as does the Yang-Mills
contribution (\ref{eq:013}) $\Bigr]$.

Although $\Pi_{ij}(p) \biggr|_{p^2 = -\mu^2}$ is finite, it is complex
and the gap equation has no solution for real $\mu^2$, owing to
unprotected threshhold singularities at $p^2 = 0$, which lead to a
complex $\Pi_{ij} (p)$ for $p^2 < 0$.

It may be that the hot thermal loop-inspired completion for the mass
term (\ref{eq:7}) is uniquely privileged in avoiding complex values
for $-\mu^2 \leq p^2 \leq 0$, but we see no reason for this.\footnote{%
We note that the hot thermal loop-inspired vertex (\ref{eq:11}) is
less singular than our (\ref{eq:016}), when any of the momentum arguments
vanish.  Correspondingly $\Pi_{ij}^N(p)$ in (\ref{eq:015}) is finite
at $p^2 = 0$, in contrast to $\Pi_{ij}(p)$ which diverges at ${1\over
p^2}$.  However, we do not recognize that this variety of singularities
at $p^2 = 0$ influences reality at $p^2 = -\mu^2$; indeed the individual
graphs contributing to $\Pi_{ij}^N$ are complex at that point, owing to
non-divergent threshhold singularities at $p^2 = 0$ that cancel in the sum.}
Absent any argument for the disappearance of the threshhold at $p^2=0$,
and reality in the region $-\mu^2 \leq p^2 < 0$, we should
expect that also the hot thermal loop-inspired calculation will exhibit
such behavior beyond the 1-loop order.\footnote{%
V.P.~Nair informs us that at the two loop level, there is evidence for
$\ln (1 + {p^2 \over \mu^2})$ terms, but it is not known whether they
acquire a protective factor of $({\mu^2 \over p^2}+1)$.}

Thus until the status of threshhold singularities is clarified, the
self-consistent gap equation for a magnetic mass provides inconclusive
evidence for magnetic mass generation.  Moreover, if there exist gauge
invariant completions for the mass term, other than the hard thermal
loop-inspired one, that lead to real $\Pi_{ij}$ at $p^2 =
-\mu^2$, it is unlikely that they all would give the same $\mu$ at one
loop level, which is further reason why higher orders must be assessed.



\begin{thebibliography}{99}


\bigskip
\frenchspacing


\bibitem{ref:1}
For summary see, {\it e.g.\/} V.P.~Nair, Lectures at Mt.~Sorak
Symposium, CCNY-HEP-94-10.

\bibitem{ref:2}
V.P.~Nair, {\it Phys.~Lett.\/}~{\bf B352} (1995) 117.

\bibitem{ref:3}
G.~Alexanian and V.P.~Nair, {\it Phys.~Lett.\/}~{\bf B352} (1995) 435.

\bibitem{ref:4}
S.~Deser, R.~Jackiw and S.~Templeton, {\it Ann.~Phys.} (NY) {\bf 140}
(1982) 372, (E) {\bf 185} (1988) 406.

\nonfrenchspacing
\end{thebibliography}
\end{document}